\documentstyle[preprint,aps]{revtex}   
\begin{document}
\draft

\title{Self-organized  superlattice formation in II-VI and III-V
semiconductors}

\author{Albert-L\'aszl\'o Barab\'asi$^*$}

\address{Department of Physics, University of Notre Dame,  Notre
Dame, IN 46556.}
\date{April 25, 1996}
\maketitle

\begin{abstract}
There is extensive recent experimental evidence of spontaneous
superlattice (SL) formation in various II-VI and III-V
semiconductors. Here we propose an atomistic mechanism
responsible
for
SL formation, and derive a relation predicting the temperature,
flux
and miscut dependence of the SL layer thickness.  Moreover, the
model
explains the existence of a critical miscut angle below which no
SL
is
formed, in agreement with results on ZnSeTe, and predicts the
formation of a platelet structure for deposition onto high
symmetry
surfaces, similar to that  observed in InAsSb.
\end{abstract}

\pacs{}
\narrowtext

The design and growth of superlattices (SLs) with optimal
electronic,
magnetic and optical properties is a challenging problem, with
major
applications for the production of electronic and optical
devices.
 In this light recent
experimental evidence on spontaneous (or self-organized) SL
formation
\cite{Ahrenkiel95,Ahrenkiel95b,Norman91} opens new, still largely
unexplored ways for creating new materials with potentially
important
electronic properties.

Spontaneous SL formation has been observed in several III-V and
II-VI
materials, including InAsSb, GaAsSb, ZnSeTe and ZnCdSeTe.  In the
ZnSe$_{1-x}$Te$_{x}$ system Zn, Se and Te were {\it
simultaneously}
deposited on a vicinal GaAs(001) substrate 
\cite{Ahrenkiel95,Ahrenkiel95b}.  Transmission electron
microscopy
and
X-Ray scattering revealed spontaneous formation of modulated
composition $x$ along the growth direction, leading to a layered
structure of varying Se and Te rich regions of thickness with a
surprisingly regular period that varied between 18 and 32 $\AA$
in
various specimens.  Interestingly, no SL was observed to form for
miscuts smaller than 4 degrees.   
In
contrast to this, InAsSb grown on high symmetry [001] surfaces
resulted in anisotropically shaped interleaved platelets of two
different alloy composition \cite{Norman91}, with  thickness 
varying between 240 and 500 $\AA$. In addition to these  two well
investigated systems, natural SLs have been also observed in
GaAsSb
and ZnCdSeTe \cite{Jacek}.

It is well established that tetrahedrally-bonded semiconductor alloys
grown by epitaxy frequently exhibit spontaneous departures from a
purely random distribution of their constituents. Specifically,
formation of {\it atomic} superlattices along various crystallographic
directions and/or phase separation have been observed and studied in
practically all III-V and several II-VI systems
\cite{Zunger}. However, the self-organized superlattices discussed
here differ from atomic ordering in that it is {\it mesoscopic} in
scale, and the period of the SL is not an integer multiple of the
lattice constant.  Closest to this behavior is phase separation,
consequently this phenomena is often called {\it vertical phase
separation} \cite{Zunger}.

For device applications we must be able to control the period of the
SL.  As a first step in this direction, here we propose an atomistic
mechanism that sheds light into the dynamics of spontaneous SL
formation and allows us to predict the temperature, flux, and miscut
angle dependence of the layer thickness.  Moreover, the model can
account for the minimal miscut angle observed in the II-VI materials,
and predicts the $T$ and $F$ dependence of this angle.  In the absence
of a miscut the model predicts the development of elongated platelets,
in accord with the experimental observation for III-V materials.

The existence of a minimal miscut for ZnSeTe indicates the key role
which the steps play during growth, strongly suggesting that the
growth mode is what is refereed to as {\it step flow}, i.e.  island
nucleation on terraces is negligible.  Indeed, the nucleation of Se or
Te rich islands on the steps would eventually destroy the long-range
order. However, SL formation implies that there is a preferential
bonding of the Se(Te) atoms to the Se(Te) rich steps. Since in the
zincblende lattice there is no direct bond between the Se and Te atoms
(neither direct Se-Se or Te-Te bonds), the information about the
chemical composition has to be transmitted through stress and
stress-generated lattice distortions \cite{newman}.  Such preferred
bonding is the result of the composition and strain induced free
energy changes, that are much studied for (mostly III-V) semiconductor
materials, often leading to clustering and phase separation
\cite{Zunger,Ipanova}.

{\it Growth model.---}  We have
the following model for the  SL formation (see Fig. 1).  Zn, Se
and
Te are deposited simultaneously on a vicinal surface, where they
diffuse.  The interface grows in a step-flow mode.  The atoms
diffuse
on the surface, the Zn atom being instantaneously trapped as it
finds
two Se and/or Te bonds.  However, Te (Se) diffuses on the
surface,
seeking a step rich  in like atoms, i.e.  Se wants to attach to
the
edge of a terrace made up of mostly  Se atoms, and likewise for
Te. 
Once it
finds such a step, we consider that it sticks instantaneously (we
shall return to this assumption below).

 It is energetically most desirable for the system to grow thick
stress-free ZnSe or ZnTe layers. For this to happen the adatoms must
have a very long diffusion length to reach the distant preferred
step. However, the adatoms (Se and Te) have a finite time to diffuse
before being "buried" by the freshly deposited atoms.  Thus the period
of the SL is limited by the ability of the Se and Te atoms to reach
such preferred step sites: the further the atoms can diffuse in the
time available for diffusion, the larger is the period of the SL.

{\it SL layer thickness---} The mechanism for SL formation,
outlined
above and shown in Fig.1, can be formulated quantitatively,
allowing
us to predict the period of the SL.  If
the
SL
has a period of $N$ monolayers (ML), the {\it average} distance a
Te
atom needs to travel to reach a Te (Se) step edge is {
proportional} to the total length of $N$ steps, $d \sim N \ell$,
where $\ell$ is the length of a single step, given by
$\ell=a/\tan \alpha$,  and  $a$ is the height of a single step. 
Assuming that the atoms follow a Brownian trajectory, the average
time, $\tau$, needed for the
diffusing atom to reach the step is given by 
$d^2 \simeq  D \tau,$
where $D$ is the diffusion constant. Here we assume that Se and
Te
have the same diffusion constant, i.e. $D_{Se}=D_{Te}=D$.
If $D_{Se} \neq D_{Te}$, then the smaller diffusion constant 
determines the period of the SL.

The adatoms must attach to their preferred step edge before being
buried by the incoming adatoms. The average lifetime of an adatom
is
given by the time necessary to deposit a full monolayer of atoms,
i.e.
$\tau = 1/F,$ where $F$ is the deposition flux in ML/sec. 
Combining
these, we find
\begin{equation}
N \simeq {\tan \alpha \over a} \left({D \over F}\right)^{1/2}.
\label{N0}
\end{equation}

 It is known
from
the theory of surface diffusion \cite{Tsong} that
\begin{equation}
D = {k_B T l_{nn}^2  \over 4 \hbar}  \exp\left[- {E_d
\over k_B T}\right],
\label{dd}
\end{equation} 
where $E_d$ is the activation energy for surface diffusion of the
adatoms, $l_{nn}$ is
the nearest
neighbor distance on the surface, and  the vibration
frequency of the surface atoms  with good approximation given by
$\nu_0 = k_B T / \hbar$. Combining  (\ref{N0}) and (\ref{dd})
we obtain 
\begin{equation}
N \simeq \left({k_B l_{nn}^2 \over  4 \hbar}\right)^{1/2}
\left({T
\over
F}\right)^{1/2} {\tan \alpha \over a} \exp\left[- {E_d \over 2
k_B
T}\right],
\label{N1}
\end{equation}
which provides the temperature-flux-miscut dependence of the  SL
period. Technically (\ref{N1}) provides the {\it largest} SL period
allowed by diffusion. However,  to decrease  its strain energy the
system wants to grow as thick as possible  ZnSe or ZnTe layers, 
this tendency being limited only by diffusion, making the largest
allowed period, (\ref{N1}),  the actual period of the SL.

The attachment of the atoms to like steps is a probabilistic
process, thus there is a nonzero probability that Se atoms would
attach to a ZnTe step, even though such bonding is energetically
less
favorable than bonding to the ZnSe step.  This makes the
transition
observed between the Se and Te rich region blurred, allowing for
a
gradual change in the composition. Indeed, X-ray diffraction data
on
the ZnSeTe system indicates the presence of a sinusoidal profile
along
the growth direction \cite{Ahrenkiel95}.

{\it Critical miscut angle.---}    The experimentally observed 
lower cutoff in the miscut angle  (critical miscut)
\cite{Ahrenkiel95} follows naturally from the model of Fig. 1:
at small miscuts $\ell$ is large and the adatoms do not have the
necessary time to reach the edge of the terrace before they are
clamped by the arrival of new atoms.  In this case we witness
{\it
island} nucleation at the surface of the terrace.  Since in the
vicinity of a ZnSe  step   Se is  captured by the
step,   most likely a ZnTe island is
 nucleated, ending the long range order.  Thus the critical
miscut
angle is the smallest miscut for which island formation is still
inhibited.   

 To evaluate the temperature and flux dependence of this
smallest
miscut  we  need to connect the typical length
scale
in
the system (for instance, the typical distance between the
islands)
to
the flux as if were no steps on the surface. 
To make progress, we use the
properties
of random walks on a plane, assuming that only monomers are
mobile.  This question has been addressed in the context of
submonolayer epitaxy, providing  the characteristic
length scale as \cite{Ghaisas92,Barabasi95}
\begin{equation}
\ell_d \sim \left({D\over F}\right)^{\psi_d},
\label{sub-4}
\end{equation}
with $\psi_d=1/6$.  
In deriving $\psi_d=1/6$  it is assumed the dimers are stable and
that the generated islands are not fractal. For extensions of
this results to other cases, including the possibility of a nonzero
critical nucleus, see
\protect\cite{Ghaisas92,Barabasi95}.  

When the typical distance between the islands, $\ell_d$, is
smaller
than the terrace size, $\ell$, island formation is observed on the top
of
the
terraces, destroying the long-range order which characterizes the
SL.
When $\ell_d >\ell$, the atoms are captured by the edge of the
steps,
and no island nucleation is expected.  The condition for island
formation,  $\ell_d  < \ell$, 
leads then  to the critical miscut angle
\begin{equation}
\tan \alpha_c \simeq \left({F \over T}\right)^{\psi_d} \exp
\left[ {
\psi_d E_d
\over k_B T}\right].
\label{alc}
\end{equation}

{\it Comparison with experiments.---} We can use Eqs. (\ref{N1}) and
(\ref{alc}) to compare experimental values of the period $N$ measured
at different miscuts, temperatures, and fluxes. The only unknown is
the diffusion energy, that we take to be $E_d= 0.5$ eV.  As a first
application we calculate the expected variation in the period of
ordering as the {\it flux and miscut are kept fixed} and the {\it
temperature is varied} between the experimentally used values, 275 and
350 $^o$C.  From Eq. (\ref{N1}) we find
${N(T_2) / N(T_1) } = 2.016,$
i.e.,  increasing the temperature  between these limits doubles
the
period of the SL.
If we keep the {\it temperature and miscut  constant}  and {\it 
increase the
flux}
 from $F_1 = 2.5$ $\AA$/s   to $F_2 = 3.5$
$\AA$/s,
the
period should  decrease by  a factor 
${N(F_2) /  N(F_1) } =  ({F_1 / F_2})^{1/2} =  0.84.$
  While the precise  temperature and
flux dependence of the SL modulation periods  in not known  yet, we
can compare these
predictions
with the reported variation in the period as the experimental
parameters were varied.  Between the mentioned temperature and
flux
limits the experimentally measured layer thickness varied between
18
$\AA$ and 32 $\AA$ \cite{Ahrenkiel95}, i.e. a factor of 1.77, in
good quantitative
agreement with the previous predictions. 

We can also calculate the expected variation in the critical 
miscut angle as the  growth parameters are varied. For the
temperature range discussed above, we find  
${ \tan \alpha_c(T_2) / \tan \alpha_c(T_1) } =  0.79,$
while for  the experimental   flux range we obtain
${\tan \alpha_c(F_2) / \tan \alpha_c(F_1) } = \left({F_2 /
F_1}\right)^{1/6} = 1.057,$
i.e.  for the temperature and flux
ranges
used during the growth, $\alpha_c$ varies only slightly (20\%
within the temperature window, and 5\% with the flux).  This
explains the experimentally observed stability of $\alpha_c$:
 the previous predictions
indicate that under the growth conditions used $\alpha_c$ is
practically independent of the growth parameters.

{\it Platelet formation and SLs in III-V
semiconductors. \cite{Norman91}}--- The previous discussion
underlies
the importance of the miscut in self-organized SL formation.  In
the
absence of a miscut the stabilizing effect of the steps is
absent,
and
the system cannot grow in a step-flow mode. Thus (\ref{N1}) is no
longer valid. In this case, one expects island formation on the
surface. However, if the same stress-induced affinity exists for
As
to
attach to InAs islands and Sb for InSb islands, this would
results
in
the segregation of InSb and InAs elongated islands, or platelets,
with
typical horizontal size $\ell_d$.  Indeed, detailed evidence
about
such platelets is presented for InAsSb grown on high symmetry
surfaces \cite{Norman91}.
 Moreover, the anisotropic nature of the platelets (they
have
different sizes along the [110] and [$\bar 1$10] directions)
indicate
the anisotropy in the energy barriers ($E_d$) for adatom
diffusion
along the two principal surface directions. Also, in the light of
the
discussed mechanism, SL formation may be induced by the
development
of
local slopes on the surface as a result of the kinetic roughening
of
the growing surface \cite{Barabasi95}. Whether such slopes
facilitate
the development of the superlattices in InAsSb is an open
question. However, for ZnSeTe, even under growth with zero
miscut,
small domains of self-organized SLs have been observed
\cite{Jacek},
most likely being correlated with the roughening of the surface,
and
the resulting local slopes with inclination larger than
$\alpha_c$.

     In conclusion, we presented a mechanism for spontaneous SL
formation, that allows us to explain both the SL observed in the II-VI
materials, and the platelets observed in the III-V systems. For growth
on vicinal surfaces we can also predict the dependence of the SL
period on the experimental parameters, that can be directly tested by
further experimental work, and can guide the growth of the SLs for
potential device applications. Furthermore, the model can explain the
origin of the lower cutoff in the miscut angle, and its relative
stability within the used experimental conditions. The understanding
of the mechanism responsible for SL formation may also help in
identifying other materials for which self-organized SL formation is
possible.


We wish to acknowledge many insightful discussions and comments by
J. Furdyna, who also generously shared with us
the
bulk of his unpublished experimental data. We have also greatly
benefited from discussions with K. Newman.

\begin{figure}
\caption{The mechanism of superlattice formation in the ZnSeTe
system.
The Te atoms (grey) want to attach to the edge of Te-rich steps,
and
Se (white) atoms attach to the edge of the Se-rich steps. Thus
the
deposited adatoms need to travel a certain distance before they
are
able to find the step edge where they can attach. The Te atom (A)
can
either aggregate with other Te atoms and nucleate a new Te island
on
the Se step, thus destroying the long range order in the system,
or
it
can diffuse (continuous line) until it reaches a Te step edge,
where
it sticks.  For SL formation the diffusion length of the Te atom
has
to be larger than the average distance it needs to travel to
reach
a
Te step.}
\label{fig1}
\end{figure}

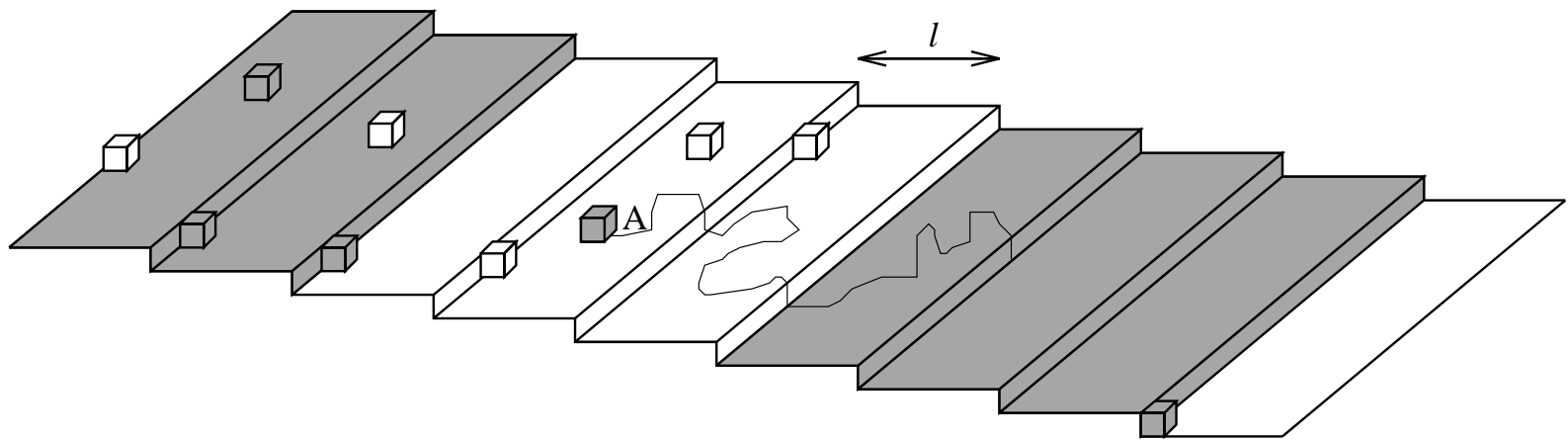
\end{document}